\documentclass[sigconf]{acmart}
\pdfoutput=1

\copyrightyear{2020}
\acmYear{2020}
\setcopyright{acmlicensed}
\acmConference[MM '20]{Proceedings of the 28th ACM International Conference on Multimedia}{October 12--16, 2020}{Seattle, WA, USA}
\acmBooktitle{Proceedings of the 28th ACM International Conference on Multimedia (MM '20), October 12--16, 2020, Seattle, WA, USA}
\acmPrice{15.00}
\acmDOI{10.1145/3394171.3413697}
\acmISBN{978-1-4503-7988-5/20/10}

\AtBeginDocument{%
  \providecommand\BibTeX{{%
    \normalfont B\kern-0.5em{\scshape i\kern-0.25em b}\kern-0.8em\TeX}}}

\usepackage{subcaption}

\begin{document}
\fancyhead{}
\title{A Modular Approach for Synchronized Wireless Multimodal Multisensor Data Acquisition in Highly Dynamic Social Settings}


\author{Chirag Raman}
\authornote{Both authors contributed equally to this research.}
\email{c.a.raman@tudelft.nl}
\affiliation{%
  \institution{Delft University of Technology}
}

\author{Stephanie Tan}
\authornotemark[1]
\email{s.tan-1@tudelft.nl}
\affiliation{%
  \institution{Delft University of Technology}
}

\author{Hayley Hung}
\email{h.hung@tudelft.nl}
\affiliation{%
  \institution{Delft University of Technology}
}


\begin{abstract}
 Existing data acquisition literature for human behavior research provides wired solutions, mainly for controlled laboratory setups. In uncontrolled free-standing conversation settings, where participants are free to walk around, these solutions are unsuitable. While wireless solutions are employed in the broadcasting industry, they can be prohibitively expensive. In this work, we propose a modular and cost-effective wireless approach for synchronized multisensor data acquisition of social human behavior. Our core idea involves a cost-accuracy trade-off by using Network Time Protocol (NTP) as a source reference for all sensors. While commonly used as a reference in ubiquitous computing, NTP is widely considered to be insufficiently accurate as a reference for video applications, where Precision Time Protocol (PTP) or Global Positioning System (GPS) based references are preferred. We argue and show, however, that the latency introduced by using NTP as a source reference is adequate for human behavior research, and the subsequent cost and modularity benefits are a desirable trade-off for applications in this domain. We also describe one instantiation of the approach deployed in a real-world experiment to demonstrate the practicality of our setup \textit{in-the-wild}.
\end{abstract}




\maketitle

\section{Introduction}

Human social behavior is a dynamic multimodal phenomenon; we express ourselves visually, vocally, and verbally.
A significant focus of research here is the complex interpersonal dynamics between interaction partners, such as turn-taking in conversations \cite{heldner2010pauses, levitan2011entrainment}, or synchrony between participants \cite{delaherche2012interpersonal}. An essential characteristic of these phenomena is their highly dynamic and multimodal nature; they evolve on  short time-scales, requiring precise synchronization of multimodal and sometimes also multisensor data streams.

\begin{figure}[!t]
    \centering
    \includegraphics[width=\columnwidth]{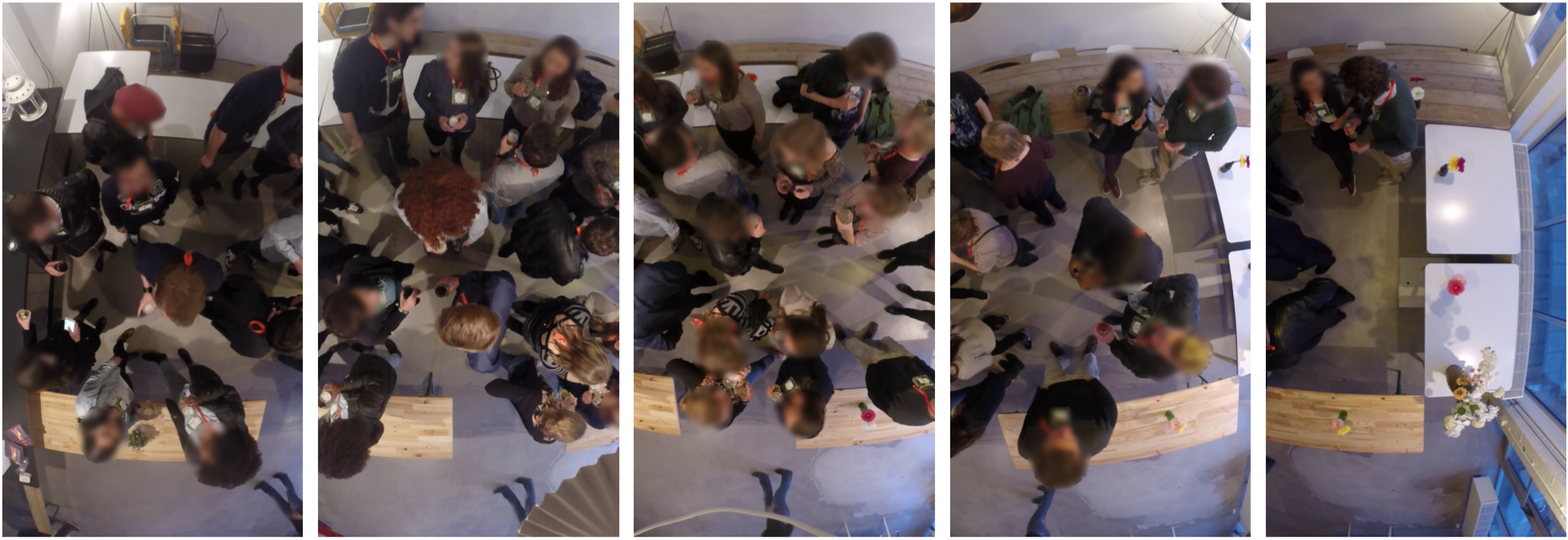}
    \caption{A typical in-the-wild social interaction setting; adapted from the MatchNMingle Dataset \cite{cabrera2018matchnmingle}}
    \label{mnm}
\end{figure}

Historically, human social behavior for automated analysis has been captured in controlled lab settings. As multimodal data analysis has become more prevalent, recorded sensors would be physically connected to relay timing information to ensure packet synchronization \cite{cao2008flexible, svoboda2002viroom, lichtenauer2011cost}. Concurrently, the ubiquitous computing community were developing approaches using wearable sensors that allowed for more pervasive sensing of social behaviors \cite{choudhury2003sensing,wyatt2010measuring,olguin2008sensible} while loosening strong requirements for data synchronization. As the trend moved towards more \textit{in-the-wild} behavior analysis, multimedia researchers turned to collecting data in more uncontrolled settings that better matched real-world scenarios. Here, multiple visual and wearable sensing sources from both modalities have been combined \cite{alameda2015salsa,cabrera2018matchnmingle}. Figure \ref{mnm} depicts a typical in-the-wild social interaction. In such prior works however, frame level synchronization requirements were circumvented by designing automated analysis approaches that smoothed behavioral data over broader time intervals on the order of a few seconds.
On the other hand, the ubiquitous computing approach has somewhat waived the need for more robust synchronization by adapting to problems that are able to take the wearable sensor data at face value and aggregate over sufficiently long time periods. This makes fine grained timing errors on the shorter scale of minutes or seconds less relevant \cite{olguin2008sensible}.

In this paper, we argue that developing any approach to analyze the fine temporal dynamics of multi-modal multi-sensor behavioral data requires us to ensure a maximum temporal latency at the data collection stage of 40~ms (see Sec. \ref{sec:human-latency} for further discussion). This requires us to bridge two traditions related to synchronization from the multimedia and ubiquitous computing domain which utilize different timing protocols and formats. Modalities such as audio and video, which have been used to analyze human behaviour analagous to human perception have used protocols such as PTP or GPS based reference time which enables sub-frame level synchronization using specialized hardware. Data here is often timestamped in the frame-based SMPTE timecode format such as linear time code (LTC)- HH:MM:SS:FF \cite{smpte12m}. Meanwhile, in the ubiquitous computing domain, sensing devices have been born out of a tradition of wireless and distributed computing where each sensing device is itself also a microcomputer and as such has used NTP \cite{mills2010network}, relying on local UNIX system time to timestamp data. While it is widely understood that PTP or GPS based timing affords superior accuracy compared to NTP, setting up a multimodal multisensor system using the specialized hardware is prohibitively expensive.


In summary, we seek to answer the following question: how can we design a modular, cost-effective, distributed multi-sensor data acquisition setup for synchronized capture of social human behaviour in-the-wild? Concretely, our contributions are as follows:

\begin{itemize}
    \item We propose and deploy a novel distributed data acquisition architecture built upon commercially available off-the-shelf components to wirelessly synchronize cameras (video) and wearable sensors (audio, inertial motion data, proximity) in-the-wild. Our core idea involves utilizing the Network Time Protocol (NTP)  \cite{mills1991internet} as a common reference for all modalities, a choice contrary to conventional use in broadcasting setups.
    \item We show that the reduced accuracy of NTP in favor of significant cost and modularity benefits is a desirable trade-off for achieving crossmodal synchronization in data recording for human behavior research applications.
\end{itemize}

\begin{figure}[!t]
    \centering
    \includegraphics[width=\columnwidth]{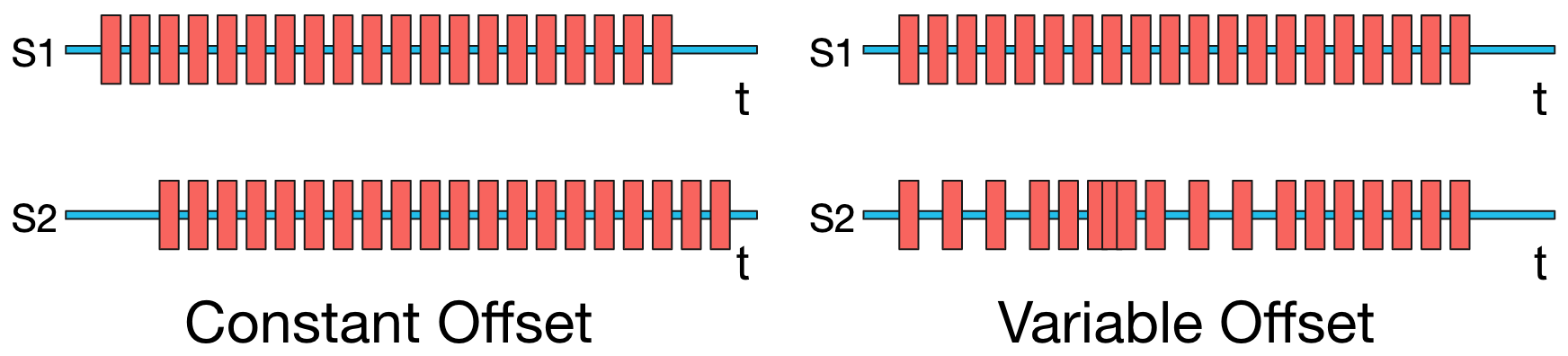}
    \caption{Basic types of desynchronization}
    \label{desync}
\end{figure}

We support our argument in the rest of this work as follows. In Section \ref{sec:related-work} we review data recording or post-processing techniques used in other human behavior research and discuss the trade-offs involved. In Section \ref{sec:approach} we establish acceptable latency tolerances for our application domain and propose our architecture, also describing a real-world instantiation of our system. We provide experiments to quantify the latency involved in our setup in Section \ref{sec:experiments} before discussing cost versus latency considerations in Section \ref{sec:discussion}. Finally, we summarize our findings in Section \ref{sec:conclusion}.

\section{Related work} \label{sec:related-work}

\textbf{Synchronization Issues.} We begin by first concretely describing the synchronization issues we propose to solve. We break these down into two basic types\textemdash constant and variable offset between data packets. Figure \ref{desync} depicts these issues for two data streams $S1$ and $S2$ over a world clock time axis $t$.

In the first case, all packets in $S2$ are offset from the corresponding packets in $S1$ by a uniform constant offset. This could arise because the triggers for recording the two streams are delayed, or because the internal clocks of the devices don't match.
In the second case, while some packets are aligned in both streams, other packets are out of sync by a variable offset, and are said to have drifted. One such common scenario involves devices recording with variable framerate or dropped packets; for instance, while recording a long session with a standard webcam with autofocus or variable framerate, the video often drifts with respect to the audio over time. In practice, both these issues occur simultaneously, and information about the world clock is required to correct for these issues directly.

\begin{figure*}[ht!]
    \centering
    \includegraphics[width=\textwidth]{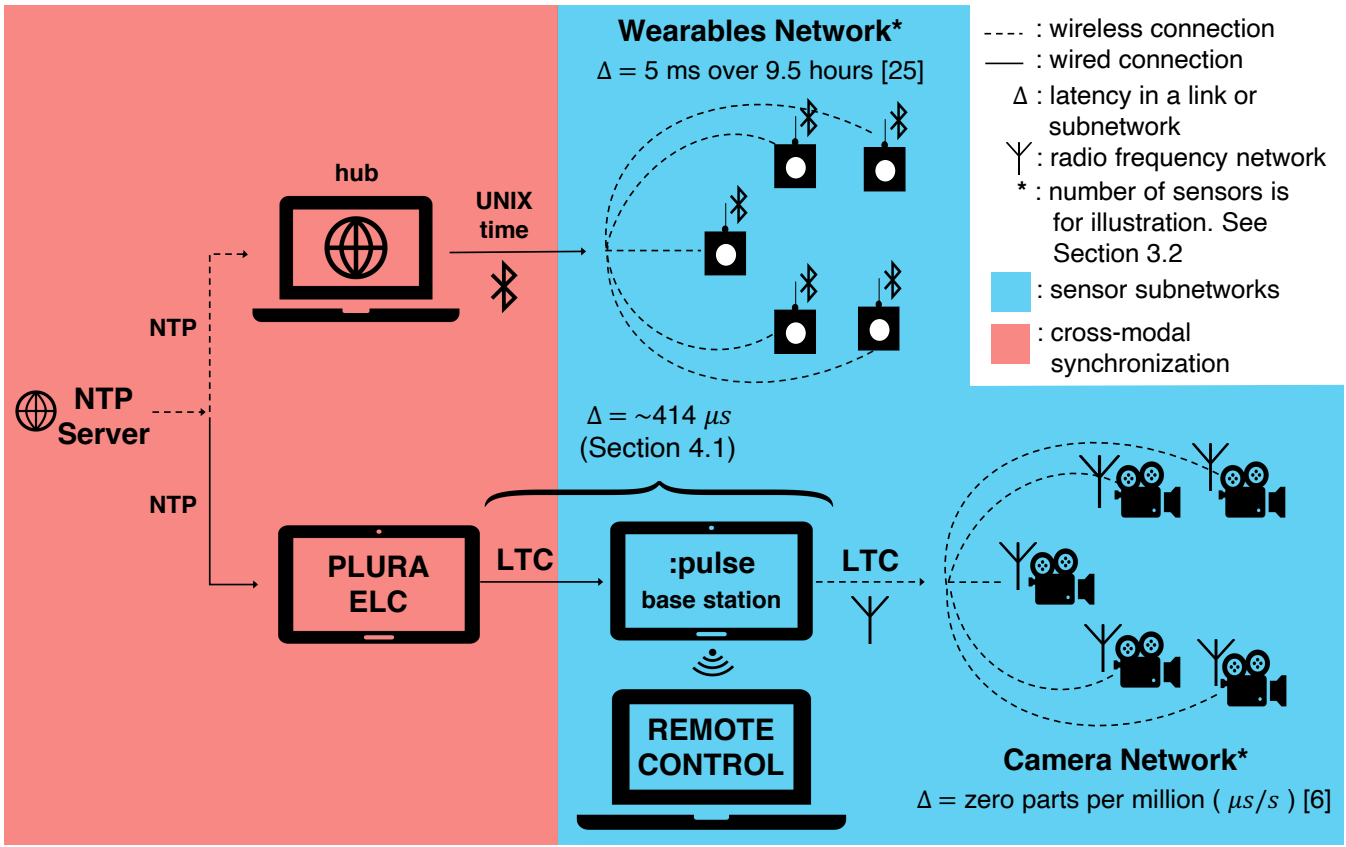}
    \caption{Overview of our proposed architecture. The reference time signal originates from the chosen NTP server and propagates to the subnetworks of wearable sensors and cameras. }
    \label{sensorNetwork}
\end{figure*}

\textbf{Event-based Approaches and Post Processing.} Many widely used human behavior datasets attempt to fix the constant offset issues in post-processing by maximizing similarity scores around a manually identified common event in data streams. Traditionally, such an event included a balloon pop, a clap or the turning off of lights to get a common dark frame across cameras. More recently, \citeauthor{alameda2015salsa} use infra-red detections in cameras and wearable sensors to compute the optimal shift according to a similarity score \cite{alameda2015salsa}.  \citeauthor{recola} use a common speech event such as the rise of a plosive to manually align high-quality audio from an external microphone to the low-quality audio from a webcam before computing the inter-correlation score around the located event \cite{recola}.
While this approach helps with fixing mismatches around a single manually identified event, they are insufficient for fixing streams that have drifted over time or have variable offset (\citeauthor{alameda2015salsa} work with a no-drift assumption). More sophisticated approaches attempt to automatically identify events for synchronizing larger parts of the streams \cite{bannach2009automatic}. In contrast, we propose a modular approach that synchronizes the devices at data acquisition, requiring minimal\textemdash if any\textemdash post processing for synchronization.

\textbf{Downstream Tasks.} In addition to fixing synchronization issues in post-processing, a common approach is to mitigate their effect on downstream tasks. The core idea is to compute features over a window \cite{gedik2018detecting, raman2019towards, rosatelli2019detecting, matic2012analysis}. The size of this window is chosen to be larger than the duration by which the modalities are assumed to be out of synchronization. The features are computed using summary statistics, or by passing the individual features through a recurrent neural network and using the last hidden state as a representation of the window.  This choice of window size, and whether this has a detrimental effect on the study of the phenomenon of interest can be contextualized by the discussion in Section \ref{sec:human-latency}.

\textbf{Ubiquitous Computing Approaches.} The analysis of social interactions has also been of interest to the ubiquitous computing community. Early work involved the development of custom wearable sensors like the UbER-Badge \cite{laibowitz2006sensor} to analyze interest and affiliation in conference attendees \cite{gips2006mapping}. Period timestamps in these setups were relayed across a Radio Frequency (RF) network every 15 minutes. \citeauthor{cattuto2010dynamics} analyzed interactions in crowded social settings using custom RFID (Radio Frequency Identification) tags \cite{cattuto2010dynamics}. Packets from the tags were relayed to radio receivers that passed it to a central server for timestamping and storage. Their approach does not record timestamp at tag acquisition, and does not account for potential delays in transmission. For modeling longitudinal social interaction networks in-the-wild, \cite{wyatt2010measuring} used personal digital assistant (PDA) devices, and found the PDAs' clocks to be "shockingly unreliable", drifting up to 5 minutes across three weeks. \citeauthor{matic2012analysis} infer interpersonal distance and relative orientation averaged over 10~s windows from up to five mobile phones in interactions lasting up to 15 minutes \cite{matic2012analysis}. They state the mobile phones had synchronized clocks without specifying how they were synchronized.

\textbf{Synchronization at Acquisition.} A significantly more accurate, albeit expensive, approach compared to those discussed involves performing synchronization at data acquisition. This is achieved at the hardware level using either software or hardware triggers. Early approaches involved connecting low-cost cameras to standard computers over an Ethernet network and using software triggers to drive the recording \cite{svoboda2002viroom, cao2008flexible}. While the cost of sensors in these setups is low, the cost of computers remains. Timing control can be improved by using a common clock and physical hardware trigger lines into the cameras in an array \cite{aha-camera}, although this only works for the video modality.

\citeauthor{lichtenauer2011cost} significantly improved over previous works by proposing a system for multimodal data capture that centralizes the synchronization task by physically connecting the sensors to a multi-channel audio interface \cite{lichtenauer2011cost}. This approach was used in the recording of the MAHNOB-HCI datasets \cite{mahnob-hci}. Other approaches have been proposed for setups involving motion-capture systems, where synchronization is achieved by plugging the output of the motion capture system to a robot in a human-robot interaction study \cite{fotinea2014data}, or in post-processing by performing an optimization over or manually annotated markers in a subset of frames \cite{sigal2010humaneva}.
These solutions are hard to deploy within in-the-wild settings over large physical areas since they are mainly wired solutions. They entail physically running trigger lines to the sensors of connecting the sensors or multiple PCs to a central audio interface.  Comparatively, our solution affords for seamless decentralized addition of sensors to the system as long as those sensors are synchronizing clocks to the common NTP reference.

The closest work matching the scale and design requirements of our interaction setup is the MatchNMingle dataset \cite{cabrera2018matchnmingle}, involving speed-dates followed by a mingling event. Their setup for the mingling event involves nine overhead GoPro cameras and wearable sensors on about 30 participants for each of three days.
GoPro cameras in their setup are triggered using an infrared remote which might induce trigger delays, and no explicit timecode synchronization is done between the cameras which each record local time. The wearable sensors are synchronized intramodally to a global timestamp accurate to 1 second \cite{dobson2013low}. The video data is synchronized manually to the wearable sensors by using a GoPro to visually record the global timestamp propagating through the wearable network displayed on a screen. In contrast, our solution achieves timecode sync at acquisition at the microsecond level for the camera network and at the millisecond level across modalities.

To the best of our knowledge, the system we propose here is the first complete distributed and scalable multi-sensor data capture solution providing timecode synchronization between modalities at data acquisition for human behavior research.

\section{Our Approach} \label{sec:approach}
\begin{figure*}
\begin{subfigure}{\columnwidth}
  \centering
  \includegraphics[width=\columnwidth]{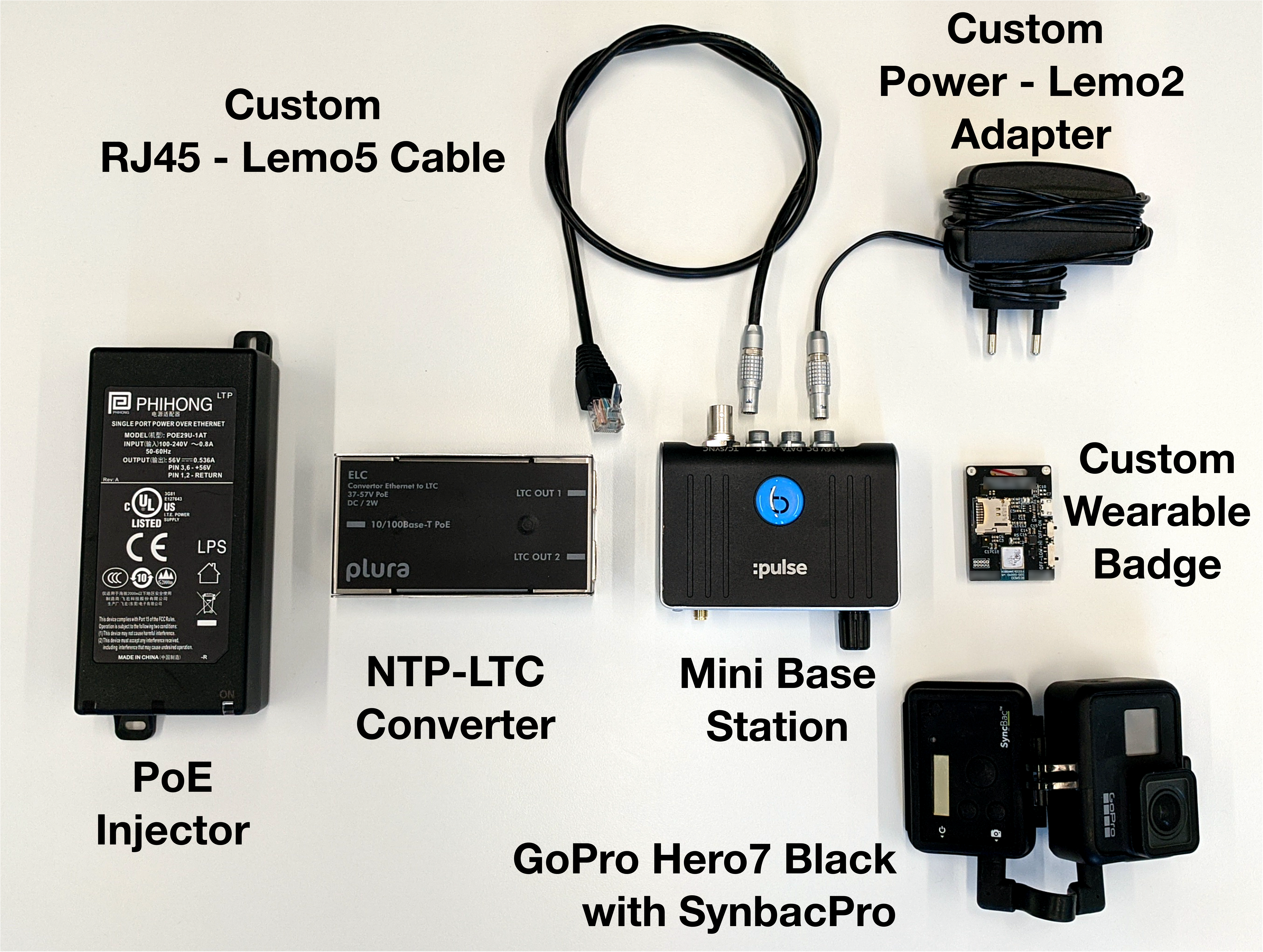}
  \caption{Core components in our setup depicted only with custom cables; the connectors are aligned with the corresponding sockets.}
  \label{fig:core-hardware}
\end{subfigure}
\begin{subfigure}{0.92\columnwidth}
  \centering
  \includegraphics[width=\columnwidth]{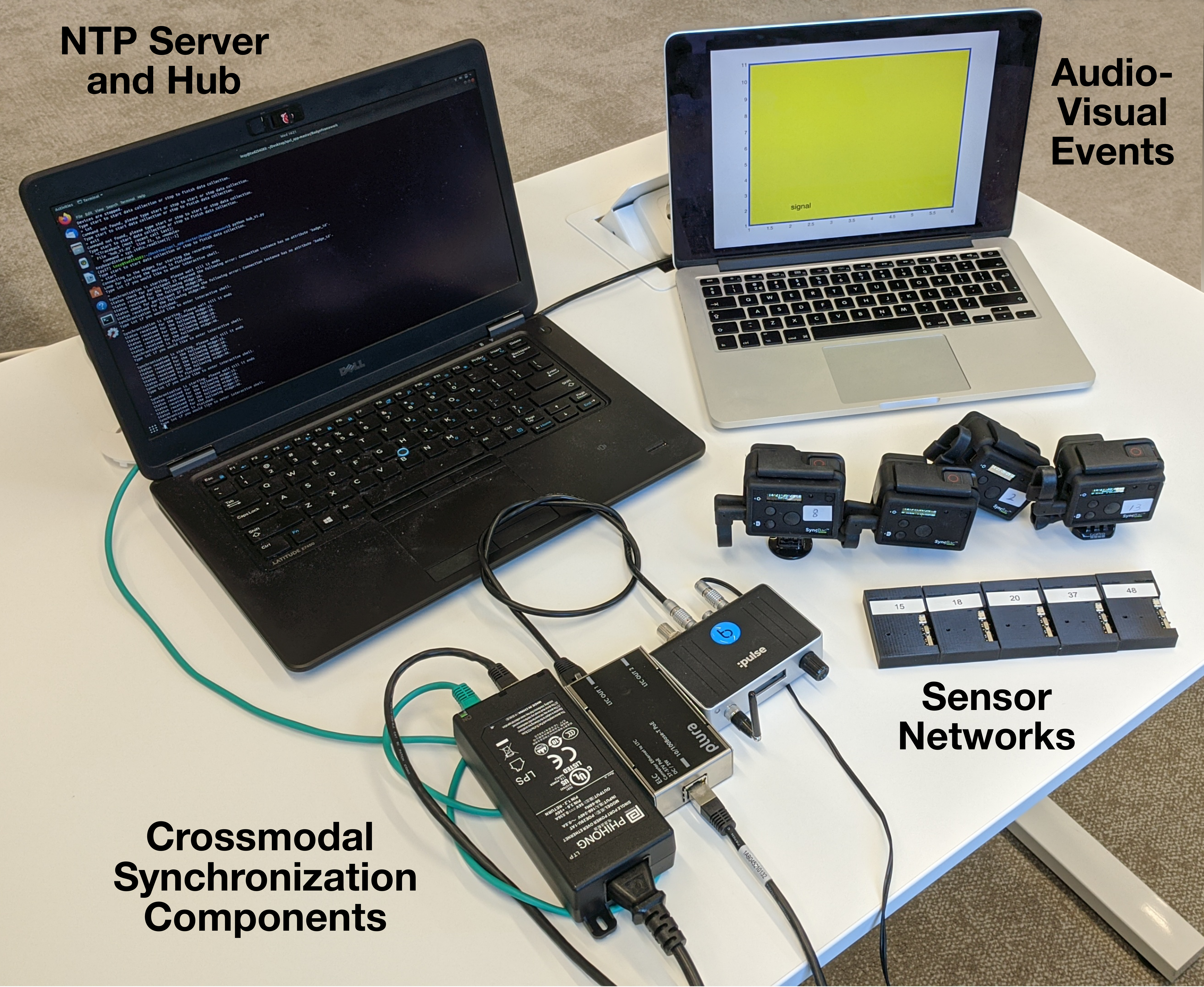}
  \caption{Full working setup of our data acquisition system, here shown with four cameras and five wearable sensors.}
  \label{fig:full-setup}
\end{subfigure}
\caption{Real-world implementation of our proposed approach. Our working setup in Figure \ref{fig:full-setup} is shown here recording audio-visual events for evaluating crossmodal synchronization, as discussed in Section \ref{subsec:crossmodal}.}
\label{fig:real-world}
\end{figure*}

Our core idea is to propagate a common time reference NTP signal to end devices (i.e., wearable sensors and cameras) at the time of data acquisition. Our approach is illustrated in Figure \ref{sensorNetwork}. The key challenge is that different subnetworks employ different timing information. The cameras use LTC for correct color framing and clock synchronization; the wearable sensors use the UNIX time received from the hub. With simply one additional hardware component (Plura ELC) combined with our choice of a common NTP reference, we achieve seamless crossmodal synchronization while preserving the existing local scheme of timekeeping. Starting from the origin of our system which is the NTP server, we explain the trade-offs of using NTP in Section \ref{sec:NTP}. We describe a particular real-world instantiation of our system in Section \ref{sec:acquisition}, where we provide implementation details on how to relay time information to the sensor subnetworks. We contextualize latency measures within the human behavior research domain in Section \ref{sec:human-latency}, which frames our subsequent experimental design.

\subsection{NTP as a reference signal} \label{sec:NTP}
The main consideration of our approach is whether using NTP as a reference for cameras recording audiovisual data compromises the latency tolerance margins of the application when compared to more commonly used higher accuracy references such as PTP and GPS. Concretely, NTP is a software based protocol. While it uses a standardized, 64-bit UDP packet that can theoretically achieve picosecond timing, the latency error for NTP is heavily dependent on the network and ambient characteristics, and is typically measured on the order of milliseconds. On the other hand, PTP (specified in the IEEE 1588 standard) utilizes hardware based timestamping \cite{eidson2002ieee} to improve over NTP latency accuracy. With customized hardware, the latency error of PTP can be guaranteed to be on the order of microseconds. Though not as accurate as PTP or GPS-based solutions, using NTP has three advantages: firstly, \textit{ease of setup}; synchronizing the system clock of a device to a local or public NTP server is straightforward, secondly, \textit{modularity}; an entire subsystem of devices can be seamlessly added to the setup and guaranteed to be synchronized with all other devices if they synchronize to a common NTP reference, and thirdly, \textit{reduced cost}; we discuss details in Section \ref{sec:discussion}. For human behavior research applications, the lowered precision trade-off in favor of increased modularity of our setup is preferable, as we further contextualize in Section \ref{sec:human-latency}.

Specifically, the clock disciplining algorithm at the heart of the NTP specification states that if left running continuously, an NTP client on a fast local area network in a home or office environment can maintain synchronization nominally within one millisecond \cite{mills_2014_discipline}. As an implementation detail, practitioners can choose between a public server such as $time.google.com$, or an isolated local NTP server at the source. Using a local server avoids upstream latency introduced by network congestion. However, using a public server provides easier setup.


\subsection{Real-world implementation} \label{sec:acquisition}

We now describe one implementation of our approach. This setup was deployed to record data from a real-world social event. It involved 48 participants each wearing a sensor around their neck, in an interaction area of size 12m x 6m, captured by elevated and overhead cameras. Our setup included the following sensors:

\begin{itemize}
    \item 13 GoPro Hero 7 Black video cameras (60fps, 1080p, Linear, NTSC) with audio (48~kHz); commercially available \cite{gopro7}.
    \item 48 custom wearable sensors adapted from the open source Rhythm Badges \cite{LedermanEtAl2017}; each sensor includes an inertial measurement unit (IMU), mono microphone (1.2~kHz), and a Bluetooth proximity sensor.
\end{itemize}

 The core components, custom hardware, and a working setup of our solution is depicted in Figure \ref{fig:real-world}. Note that in keeping with privacy regulations, the wearable sensors record audio at frequencies only sufficient for detecting voice activity rather than verbal content. This makes the already subjective task of identifying semantic event boundaries in-the-wild even harder. Consequently, for the post-hoc evaluation of our system and comparison against widely used approaches in the domain that rely on such events for synchronization, we take a more principled approach to defining and sampling stimulus events, as we discuss in Section \ref{sec:experiments}. While the number of devices we report here were used in our real-world deployment, it is not the system limit, as we discuss below. Our system is modular and scalable to larger number of devices with additional hubs and base stations (indicated in Figure \ref{sensorNetwork}).

\textbf{Relaying time to cameras.} We explain the bottom branch in Figure \ref{sensorNetwork} regarding the camera network and its upstream components in this section. A laptop that receives the time reference from a local NTP server (same as the one used by the Bluetooth hub) shares the network time through a Power-Over-Ethernet injector (Plura 30W Single Port) with an Ethernet-to-LTC Converter (Plura ELC) \cite{EthernettoLTCConvertor}. The LTC signal that is converted from NTP is sent to a base station unit by Timecode Systems called :pulse \cite{pulse}, which allows for control, synchronization and metadata exchange for all devices within the camera network. It serves as the master in the localized master-slave radio frequency (RF) network, which shares its timecode with slave devices called Syncbac PRO \cite{syncbac}, also manufactured by Timecode Systems. Each Syncbac PRO is physically tethered to a GoPro camera so that the accurate shared timecode is embedded within the MP4 files in each camera. In practice, once the timecode information of each video is available, any common video editing software can be used to align the video streams automatically for playback. An important consideration of our system design is to start the data acquisition remotely and wirelessly, since cameras are often mounted on the ceiling or other inaccessible places. The BLINK Hub app is used to remotely control (e.g. start, stop, etc), monitor and set features of all units within the localized RF network, which includes :pulse and Syncbac PRO. The BLINK Hub app can control up to 64 devices over a range of 500 m line of sight. Each :pulse unit can theoretically connect to an unlimited number of Syncbac PRO slaves within the same RF network over a range of 200 m line of sight. Both the RF network and the BLINK hub app control could have more network latency with increasing number of connections on the specific RF channel. The accuracy of the RF network synchronization is zero parts per million when the slaves (Syncbac PROs) are locked to the master (:pulse) \cite{pulse, syncbac}.

Note that our use of the ELC is different from its typical application of providing a signal for displaying the reference from a dedicated master reference generator. The novelty of our system stems from not requiring a typical GPS master reference generator at the source to phase lock to. Since our approach uses the local NTP server as the main reference itself, our use of the ELC allows for a simple method for video reference generation. Through experiments in Section \ref{sec:experiments} we show that our setup is appropriate for the domain. With the addition of a single component (any hardware or software NTP-LTC converter, the ELC in our setup), we wirelessly achieve crossmodal synchronization  between the camera and wearables network compared to previous works as well as the more expensive GPS-based setup described in Section \ref{sec:experiments}. Specifically, we are able to wirelessly embed the timecode generated from the same reference used for other subnetworks into the video files, while relying on commercial products (with only custom connecting cables) for easier reproduction.

\textbf{Relaying time to wearable sensors.} We explain the top branch in Figure \ref{sensorNetwork} regarding the wearable sensors network in this section. Note that our system design is agnostic to the choice of the type of wearable sensors. Our choice of wearable sensors for this specific instantiation is motivated by the open source platform \cite{LedermanEtAl2017} for its accessibility and reproducibility, but could be replaced by any other subnetwork of sensors\textemdash wearable or otherwise\textemdash that supports NTP time synchronization. In our system, a hub node (in form of a laptop) receives the NTP time reference and shares it with the wearable sensors. The hub connects to the sensors sequentially in order of their MAC addresses for a Bluetooth handshake that transmits the UNIX time from the hub to the sensor. Each sensor then updates its system time to this timestamp. The frequency of establishing connection (i.e., synchronization messages) is a user defined parameter, and it has been shown that any interval between 0 and 600 seconds would be appropriate \cite{hopfengaertner2018open}. Since the hub is not maintaining a connection with all sensors at all times, there is no limit on the number of sensors that the hub can connect to. In practice, the maximum number of sensors associated to the hub is dictated by the saturation of wireless channel (i.e., when collisions occur). The mean average error in synchronization within the sensor network has been shown to be 5~ms over 9.5~hours of recording \cite{hopfengaertner2018open}. While intramodal synchronization within this subnetwork can be improved through various methods such as tracking the timestamps at each timestamp reception and parallelization of communication between the hub and the sensors, such improvements are outside the scope of our contribution.

We thereby achieve multisensor intramodal synchronization, multicamera intramodal synchronization, as well as multisensor-multicamera crossmodal synchronization. To summarize, each wearable is timestamped with the UNIX system time of the wearable network hub. The hub is set to the time of the local NTP server also providing time reference to the cameras, which are then recorded in terms of LTC.
In post-processing, we convert the UNIX time to UTC time (HH:MM:SS:mS) to match samples to video frames denoted by LTC timecode (HH:MM:SS:FF). Note that these post-processing steps are insignificant compared to ones taken in manual alignment.

\subsection{Latency measures in social literature} \label{sec:human-latency}
To contextualize our assessment of tolerable latency margins, we review representative literature from social psychology that alludes to latency measures across different behavioral phenomena.

Measuring human response time (between stimulus and reaction) is an intuitive way to quantify behavior latencies. Early works have found that the response time spans between 120~ms and 300~ms \cite{luce1986response}, with a specific example finding a 157~ms latency in speech perception \cite{fowler2003rapid}. Related to speech behavior is the more complicated turn-taking mechanism in conversations that involves pauses, gaps and overlaps. The time frame of consideration in identifying gaps between speakers. (speaker change) is approximately 200~ms, which is shown to be suitable for the task \cite{heldner2010pauses}. Studies in synchrony, mimicry, entrainment, and other higher-level social phenomena usually consider a larger window size. \citeauthor{levitan2011entrainment} have shown that a window size of 200-1000~ms works well in practice for studying speech backchannels. An episode of facial and body motor mimicry could be between 40~ms and 4~s \cite{sonnby2003emotional, bilakhia2015mahnob}.

Apart from surveying the size of time frame used in various studies, an important measure of time offset is the latency in human perception of audiovisual data , since many human behavior datasets are manually annotated. Humans are shown to tolerate an audio lag of 200~ms or a video lag of 45~ms \cite{grant2003discrimination}. A successful automated method of data synchronization should perform on par with, if not better than human perception. It is worth noting that humans cannot annotate sensor data such as acceleration, in which case an automated synchronization solution is needed if aligning such data is required.

\begin{figure}[t]
    \centering
    \includegraphics[width=0.8\columnwidth]{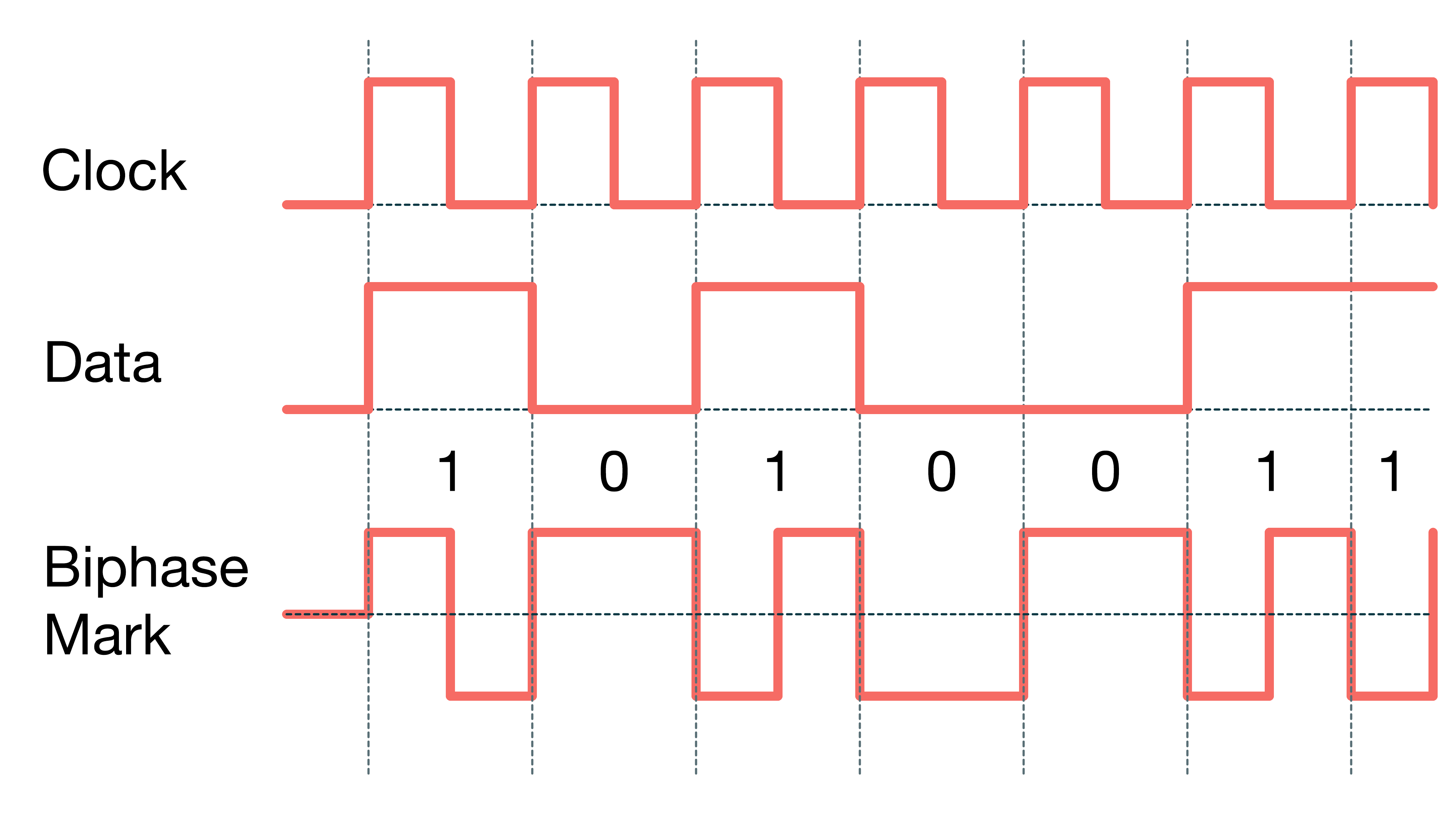}
    \caption{Biphase Mark Encoding of Linear Time Code}
    \label{ltcbiphase}
\end{figure}

We deduce that offsets within a window size and/or range of human perception error, are generally tolerable. Based on the studies listed above, we consider a time offset to be acceptable if it is between 40~ms (e.g., facial analysis) and 1000~ms (e.g., entrainment). Though smaller offsets between different data streams can be achieved, the incremental gain becomes less relevant, especially for common phenomena of interest as discussed above. Nevertheless, our setup\textemdash in which we achieve a median video latency of 414~$\mu$s and wearable data latency of 5~ms over 9.5~hours \cite{hopfengaertner2018open}\textemdash is also applicable to data collection situations where fine details like faces are important such as egocentric vision setups, or those involving physiological sensors.
\hfill \break

\section{Experiments} \label{sec:experiments}

The primary metric for synchronization accuracy is timing latency. A principled evaluation of our system would require characterizing latency at the local connection links in our proposed architecture, as well as final latency in the recorded data streams.

A common method for crossmodal synchronization used by human behavior datasets is the aligning of semantic events \cite{recola, alameda2015salsa}. As discussed in Section \ref{sec:acquisition}, given the subjective nature of start and end boundaries of semantic social events and low frequency audio recordings from wearables for privacy, we employ a more principled approach of defining and sampling stimulus ground-truth audio-visual events for our experiment presented in Section \ref{subsec:crossmodal}. Note that while the ground truth events are manually generated for control, the synchronization setup exactly matches the one we deployed in our in-the-wild experiment.

Our core crossmodal approach introduces one point of latency through the use of an NTP-LTC converter to share the common NTP reference with the camera subnetwork. Since limited hardware connections prevent recording the output LTC streams during real-world deployment, we first present a pre-experiment to measure latency at the isolated connection in Section \ref{subsec:pulse-plura}. Latency measures in our individual sensor subnetworks are depicted in Figure \ref{sensorNetwork} and already discussed in Section \ref{sec:acquisition}.

With these time drifts quantified, we demonstrate that our approach is more robust and suitable for video, audio, and wearable sensor data alignment for the purpose of studying human behavior compared to previous approaches. Code and data for the decoding and analysis in these experiments are publicly available.\footnote{Code \& data are available at https://github.com/TUDelft-SPC-Lab/sync-experiments}

\begin{figure}[!t]
    \centering
    \includegraphics[width=\columnwidth]{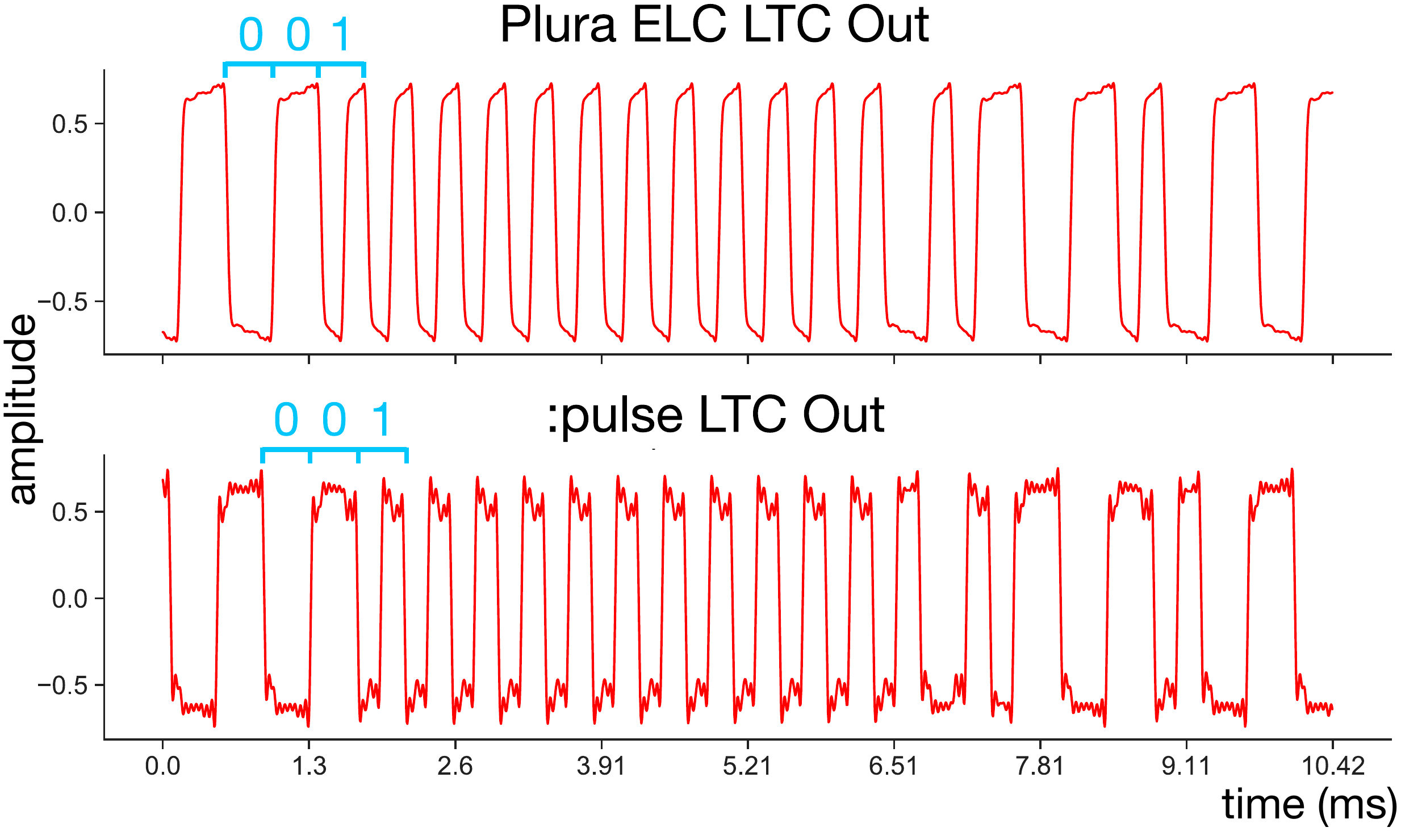}
    \caption{Raw audio LTC signals generated by the Plura ELC and :pulse modules. The window includes the encoding of an LTC sync word ($0011111111111101$) followed by the bits $0001000$ from the next frame. The lower signal here leads the upper signal by 62 audio samples, or less than 1 bit of data.}
    \label{plura-pulse}
\end{figure}

\subsection{Timecode latency between NTP-LTC converter and camera network master} \label{subsec:pulse-plura}

 We use the Plura Ethernet to LTC converter (ELC) for passing an LTC signal generated from the common NTP reference into the :pulse base station, as a timing reference for the camera network. In this experiment we evaluate the latency between two LTC signals: the LTC output of Plura ELC and the LTC output of :pulse.

\textbf{Encoding.} LTC is an encoding of timecode data within an audio signal. The timecode data is in the \textit{hour:minute:second:frame} format. The data bits in an LTC signal are encoded using the biphase mark code (BMC) as depicted in Figure \ref{ltcbiphase}: a 0 bit has a single zero-one transition at the start of the bit period; a 1 bit has two transitions, at the beginning and middle of the bit period. Each LTC frame is made up of 80 bits of data, including a 16 bits long 'sync word' $0011111111111101$ denoting the end of a frame. Consequently, at a framerate of 30~frames/sec, the LTC timecode has a maximum frequency of 2400~Hz (binary ones). In our experiments we measure the latency between the two LTC signals at the smallest possible time resolution; we consequently record the audio signals at the highest possible sampling frequency of 192~kHz, allowing for the smallest latency resolution of about 5 microseconds. Note that here theoretically, 80 audio samples correspond to 1 bit of data, and 80 bits correspond to 1 LTC frame.

\begin{figure}[b]
    \centering
    \includegraphics[width=\columnwidth]{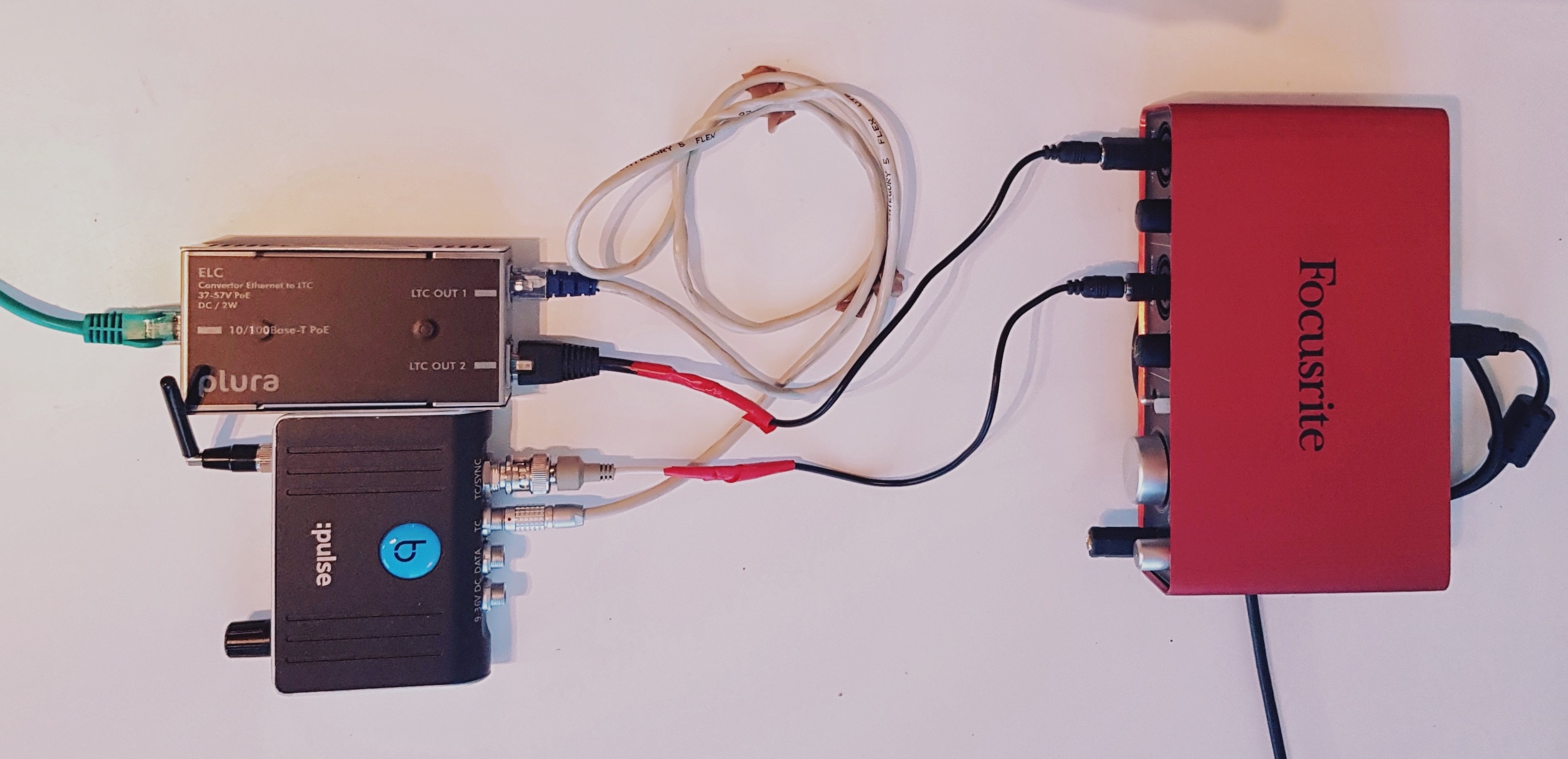}
    \caption{Hardware setup with custom cables for recording LTC signals from the Plura ELC and the :pulse base station.}
    \label{fig:plura-pulse-hardware}
\end{figure}

\textbf{Test setup and data.} We passed the outputs of the Plura ELC (RJ45 jack) and the :pulse (BNC socket) to a Focusrite Scarlett 2i2 audio interface \cite{scarlett} through custom cables. Figure \ref{fig:plura-pulse-hardware} depicts a part of our setup for recording the signals from the two devices. The Plura ELC was configured to use the public NTP server $time.google.com$ as reference and generate an LTC signal at 30 frames/second. An isolated private NTP server can also be used upstream as mentioned, but that does not affect the outcome of the latency between the ELC and the :pulse we are studying here. The LTC signals were recorded using the application Audacity. We recorded for a total duration of 30 minutes over six sessions of five minutes each, for a total of 54000 LTC frames. Figure \ref{plura-pulse} depicts a window from our recorded audio signals at the end of a frame. The signals here represent the real-world noisy LTC signals encoded using the biphase mark code depicted in Figure \ref{ltcbiphase}.

\textbf{Experiments.} We measure synchronization at two levels: LTC frame level, and audio sample level. We use \textit{demodulation} to refer to the conversion of the audio signal to binary data, and \textit{decoding} to the conversion of the binarized data into the \textit{hour:minute:second:frame} format.  The recorded audio signals have imperfect leading and falling edges along with noise, with optima corresponding to a single data bit period being between 77-83 samples apart instead of the theoretical 80 audio samples. During demodulation, we begin by finding the local optima within a window size of six samples around the 80th sample following an optima. This new optima becomes the reference for the subsequent clock period. The demodulation was verified to match the original timecode presented in the recordings on the devices. We conducted a synchronization test using the 30 minutes of recording from six sessions where the binarized stream following the first sync word was decoded into timecode for checking correspondence at the frame level. We found that the data was indeed synchronized at the frame level for all the frames. With frame-level synchronization verified, we measured the world clock latency between the signals at the sub-frame level. We do this by finding the shift in number of audio samples to achieve maximum cross-correlation between the two audio signals. This lag was found to be [79, 80, 80, 80, -43, 78] samples for our six recordings, yielding a mean latency of 307.29 microseconds (59 samples) and a median latency of 414 microseconds (79.5 samples). A positive lag implies that the :pulse signal leads the Plura ELC while a negative one implies the opposite. One way to interpret this is that the median latency is approximately 1 bit of data, which corresponds to 1/80th of an LTC frame. We conclude that this measure of latency is an order of magnitude lower than our overall acceptable latency tolerance of about 40~ms for the application domain as established in Section \ref{sec:human-latency}.

\subsection{Evaluating crossmodal synchronization} \label{subsec:crossmodal}

Assuming that the GoPro audio and video are synchronized, we compare the audio recorded by the wearable sensors with the audio recorded by the GoPros in order to evaluate crossmodal synchronization of the wearable sensors and cameras of our system. We defined 10 stimulus audio-visual events that occurred randomly based on interval length (from 1-5 seconds) sampled from a Poisson distribution. An event is comprised of a visual color change accompanied by an audio \textit{beep}. These events can be seen as the ground truth events in which the duration between each event is known. Figure \ref{fig:full-setup} depicts our full working setup for recording these events.

\begin{figure}[t]
    \centering
    \includegraphics[width=0.9\columnwidth]{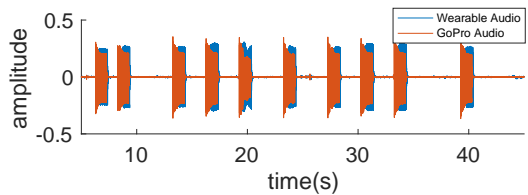}
    \caption{Representative example showing the aligned audio events in one of the wearable sensors and one of the GoPros.}
    \label{wearablesensor-gopro}
\end{figure}

The experiment considers 4 wearable sensor sensors and 4 GoPro cameras simultaneously capturing the generated audiovisual events played over approximately one minute. Figure \ref{wearablesensor-gopro} is a representative example showing that the audio events from one of the wearable sensors and one of the GoPro cameras appear to be in alignment. To further quantify the time offsets between different audio streams, we determine the number of samples between the end of an audio event and the onset of the subsequent event by thresholding the amplitude. Since the sampling frequencies of the wearable sensors (20~kHz) and the GoPros (48~kHz) are known, the number of samples is converted to time duration in seconds. We compare these empirically found durations from the recordings to ground truth durations between events .

We found that the average time offset for all wearable sensors and all GoPro recordings is 10.8$\pm$5.6~ms and 1.9$\pm$2.0~ms, respectively, when compared to the ground truth durations. Therefore, the maximum offset on average between wearable sensor and GoPro audio signals is the sum of these offsets, resulting in approximately 13~ms, for a conservative estimation. In light of the latency in upstream links which are orders of magnitude smaller than what we observe here in the end devices, we offer some hypotheses on the possible sources of errors. Firstly, there is uncertainty in the generation and transmission of synchronization messages between the hub and the wearable sensors, ranging from a few milliseconds to several seconds, depending on connection interval settings \cite{hopfengaertner2018open, bluetooth}. The time offset between the hub and the wearable sensors is inversely proportional to the frequency of connection. While it is possible to address this random time offset in Bluetooth connections via the Media Access Control (MAC) layer of the communication interface, the current approach is optimized towards energy efficiency \cite{hopfengaertner2018open}. Other possible reasons include varied quality of the wearable sensors and GoPro cameras resulting in discrepancy in sensor behavior and sensitivity, and offsets between the playback of the audiovisual events on the laptop (in Figure \ref{fig:full-setup}) and the actual recording by the sensors. Despite the 13~ms offset across the camera and wearable sensor modalities, we highlight that it is still lower than both, the lower bound of 40~ms described in Section \ref{sec:human-latency} and the human perception tolerance limit of audiovisual skew which is $\pm$ 80~ms \cite{steinmetz1996human}. In these purely perceptual tests, we could not hear any audible differences when the GoPro audio and the wearable sensor audio are played simultaneously. This shows that our approach is at least as good as, if not better than manual alignment of multimodal signals in the context of this experiment.

\section{Cost versus Latency Considerations} \label{sec:discussion}

Apart from providing a seamless interface for synchronizing different subnetworks of sensors, our choice of leveraging NTP as the common reference is also motivated by cost\textemdash the only component we have introduced to achieve crossmodal synchronization is the NTP to LTC converter. We have also shown that the reduced accuracy of our choice is well within tolerable latencies between sensors for our application domain. But what if cost is not a constraint?

For setups enjoying higher budgets, we recommend using synchronization references from highly-accurate GPS satellites.  These satellites are all synchronized to the same time using stabilized atomic clock hardware and known locations due to their medium earth orbits. As a result, GPS receivers can listen to multiple broadcast sources and use trilateration (somewhat similar to triangulation) to determine their own position and time deviation. GPS modules can consequently perform time-synchronization with a resolution of 100 nanoseconds or smaller \cite{sazonov2010wireless}.

 Through the use of satellites, a GPS based solution largely mitigates issues like unquantifiable delays in network communications or a lack of local operating system resources commonly plaguing the use of the protocols described in Section \ref{sec:NTP}. Additionally, GPS modules can be used to generate NTP and PTP signals \cite{volgyesi2017time} for downstream subnetworks. One potential downside of using GPS references is that the GPS antenna needs to be installed outdoors under visible sky to obtain the GPS reference, which might pose logistical challenges depending on the physical setting of the interactions being studied.

Since we use the Plura ELC in our setup, for comparison we provide an example GPS controlled setup using components from Plura. This involves modules from their Rubidium Series \cite{plurarubidium}. A GPS receiver such as the RUB G16X would obtain the GPS signal and pass it as reference to the RUB GT master timecode generator module to produce an LTC signal. This LTC signal would act as an external reference for the :pulse base station like in our current setup. A RUB PM-N module connected to the the GT would serve the dual purpose of powering the setup and acting as an NTP server to generate the NTP signal for the hub of the wearable sensor network similar to our current setup. The entire setup would be housed in a RUB H1 rack. The GPS setup for crossmodal synchronization is approximately eight times more expensive than our setup using an ELC and a POE injector. \footnote{The GPS setup described currently costs approximately US \${5700}, while the combined cost of the ELC and the POE injector is about US \${730}.}

\section{Conclusion} \label{sec:conclusion}

In this paper we introduce a novel approach for synchronized and wireless acquisition of human behavior data across video, audio, and wearable sensor data modalities, captured in highly dynamic in-the-wild settings. The key challenge of synchronization in these settings is to propagate a common time reference signal to end devices such as cameras and wearable sensors in a wireless and scalable manner without compounding network delays. Another challenge is that different types of sensors rely on different types of timing information. Existing solutions in this space are either wired solutions, or achieve limited synchronization in post-processing, making them less suitable for our scenario involving a large number of people free to move in a large physical area. Our novel solution uses a common NTP reference signal for both the camera and wearable sensors modalities; conventionally NTP is superceded by more accurate reference signals for video. Through empirical experiments, we show that the median time latency introduced by our choice of using NTP is 414~$\mu$s for the video modality. The intramodal latency of our wearable sensor network built by extending an open platform is 5~ms over 9.5~hours \cite{hopfengaertner2018open}. The overall crossmodal latency of our setup is approximately 13~ms at worst based on an events-based experiment. We contextualized our findings using latency measures from representative social behaviour literature, and find that our setup performs well within a tolerable latency margin of 40~ms for our application domain and human perception. To the best of our knowledge, this is the first work that quantifies latency tolerances for a data collection system designed for collecting human behavior data, and proposes a distributed architecture built on commercially available products. Through valid trade-offs, our approach provides a practical, accurate, cost-effective, time-efficient, and modular solution that is more advantageous than the current state-of-the art methods/heuristics for highly dynamic social settings.

\begin{acks}
This research was partially funded by the Netherlands Organization for
Scientific Research (NWO) under the MINGLE project number 639.022.606. We thank Ruud de Jong, Jeroen Bastemeijer, Amelia Villegas, Paul Scurrell, Thomas Rock, J\"{u}rgen Loh, and Ekin Gedik for sharing their technical expertise and giving us helpful feedback.
\end{acks}

\bibliographystyle{ACM-Reference-Format}
\bibliography{syncPaper}

\end{document}